\documentclass[a4paper,11pt]{article}
\usepackage[utf8]{inputenc}
\pdfoutput=1

\usepackage{jheppub,float,amsmath} 
\usepackage{mathtools}
\usepackage{xcolor}

\usepackage{parskip}

\usepackage{tocloft}

\usepackage{hyperref}
\usepackage{refstyle}
\usepackage{prettyref}
\newcommand{\pref}{\prettyref}
\newrefformat{app}{Appendix \ref{#1}}
\newrefformat{fig}{Figure \ref{#1}}
\usepackage{slashed}
\usepackage{mathrsfs}

\usepackage{cleveref}
\usepackage{tikz}
\usepackage[compat=1.1.0]{tikz-feynman}
\newcommand{\mc}[1]{\mathcal{#1}}
\newcommand{\ttb}{\text{T}\overline{\text{T}}}
\newcommand{\dd}{\text{d}}

\title{Renormalisation in $ \ttb $-Deformed (Non-)Integrable Theories}

\author[a]{Subhroneel Chakrabarti,}
\author[b]{Arkajyoti Manna,}
\author[c]{and Madhusudhan Raman }
\affiliation[a]{FZU - Institute of Physics of the Czech Academy of Sciences \& CEICO\\ Na Slovance 2, 182 21 Prague 8, Czech Republic}
\affiliation[b]{Institute of Mathematical Sciences, Homi Bhabha National Institute (HBNI)\\ IV Cross Road, C. I. T. Campus, Taramani, Chennai 600 113, Tamil Nadu, India}
\affiliation[c]{Instituto de Física Teórica, Universidade Estadual Paulista (IFT-UNESP)\\ R. Dr. Bento Teobaldo Ferraz 271, São Paulo 01140-070, Brazil\\}

\emailAdd{subhroneelc@fzu.cz}
\emailAdd{arkajyotim@imsc.res.in}
\emailAdd{madhusudhan.raman@unesp.br}

\abstract{That the exact quantum S-matrix of $\ttb$-deformed field
  theories is known has interesting consequences for their
  perturbative renormalisation. Recent investigations into the
  interplay between renormalisation and integrable deformations have
  focused on those cases where the (undeformed) seed theories are
  integrable. In this paper, we study the perturbative renormalisation
  of non-integrable $ \ttb $-deformed field theories, focusing on a theory of scalars
  with a quartic interaction. We show that in this theory, the
  $ 4 $-point $ 1 $-loop S-matrix exhibits features similar to its
  integrable cousins. We also carry out a similar analysis for the
  case of an integrable quantum field theory comprising a pair of
  chiral bosons. Our discussion of this latter theory supplies the first
  explicit example of a perturbative field theory computation in the
  formalism due to Sen.}

\begin{document} 
\maketitle
\flushbottom

\section{Introduction and Methodology}
Solvable irrelevant deformations of two-dimensional quantum field theories
such as the $\ttb$ deformation \citep{Zamolodchikov:2004ce,Cavaglia:2016oda,Smirnov:2016lqw} have
several novel and interesting features, that have resulted in the widespread
attention they have received over the last few years. (We refer the reader
to \cite{Jiang:2019hxb} for a comprehensive review of recent
developments in this field.) The central property of interest in this
paper is the following rather remarkable result:  the exact quantum
S-matrix $ S_{\lambda} $ of the $\ttb$-deformed theory is related to the quantum
S-matrix $ S_{0} $ of the undeformed seed
theory as
\begin{equation} \label{eq:smatrix}
  S_{\lambda} = S_{\text{CDD}} \, S_{0} \ .
\end{equation}
Here, $S_{\text{CDD}}$ is a pure phase, known as the
Castillejo-Dalitz-Dyson (CDD) factor, and it captures the effect of
$\ttb$ deformations on the S-matrix of the seed theory.

Being an exact quantum statement, \eqref{smatrix} has a profound
consequences. For instance, it implies that if we start with the
classical $\ttb$-deformed Lagrangian and renormalise it following the
usual techniques, the S-matrix obtained from the renormalised
Lagrangian must be consistent with \eqref{smatrix} order-by-order in
perturbation theory. Said differently, knowledge of the exact
deformed S-matrix imposes strong constraints on the renormalisation of
the classical deformed theory. This is a remarkable result, especially
considering the irrelevant nature of $\ttb$ deformation. 

Following this line of reasoning, it was shown in
\citep{Rosenhaus:2019utc} that a $ \ttb $-deformed free scalar, with
or without mass, can be uniquely renormalised in such a way that
amplitudes from renormalised perturbation theory match
\eqref{smatrix}. A similar success story for the case of $ \ttb $-deformed
free massive fermions was reported in \citep{Dey:2021jyl}. In this
paper, we extend these studies in two important ways.

\subsection*{Chiral and Integrable Theories}

First, we study the perturbative renormalisation of a $\ttb$-deformed
theory of left- and right-chiral bosons whose deformed classical
Lagrangian was determined and studied in
\citep{Chakrabarti:2020dhv}. This, like the previously studied cases
in the literature, is another example of an integrable quantum field
theory. However, owing to the fact that these theories can only be defined in Lorentzian signature, it occupies a rather unusual position in the bestiary
of two dimensional quantum field theories. 

Recall that the $\ttb$
deformation is only rigorously defined in Euclidean
signature. The easiest way to see this is to observe that the workhorse of this deformation --- the Zamolodchikov factorisation formula --- is derived using Euclidean operator product expansions. In deriving \eqref{smatrix} it was implicitly assumed that
there is no obstruction to analytically continuing the (un)deformed
Euclidean theory to the corresponding Lorentzian theory. While this is
ordinarily true for most quantum field theories, in the case of chiral
bosons this assumption fails. In \citep{Chakrabarti:2020dhv} it was
\emph{assumed} that the classical deformed Lagrangian can be determined using
the Lorentzian version of the $\ttb$ operator. This
requires substantiation. We use \eqref{smatrix} as the \emph{definition}
of the $\ttb$ deformation of a Lorentzian theory. It is then a
non-trivial check to see if the classical deformed Lagrangian obtained
in \citep{Chakrabarti:2020dhv} can also be renormalised such that it
matches with the CDD deformation of the S-matrix. This is one of the goals of this paper.

This example is also interesting in its own right. A truly Lorentz
covariant polynomial action for chiral bosons with chirality imposed
off-shell as well as on-shell was only discovered recently
\citep{Sen:2015nph,Sen:2019qit}. Such chiral boson actions have
several unusual features, such as inclusion of an additional free
field with wrong-signed kinetic term which completely decouples from
the dynamics, but not at the level of action. A perturbative loop-level computation of S-matrix for field theories with these unusual
actions, to the best of our knowledge, has not been done so
far.\footnote{While it was explained in \citep{Sen:2015nph} how one might
  go about determining the Feynman rules for such actions, no explicit perturbative
  calculations with this action were performed.} This computation thus
should also provide a new result in the interacting quantum theories
of chiral bosons.

\subsection*{Non-Integrable Theories}

Second, we consider the deformations of interacting non-integrable
quantum field theories and their renormalisation. In particular, we consider the case of a massive scalar with a
quartic interaction as our seed theory and consider the
perturbative renormalisation of its $ \ttb $ deformation. 

There are
two principal novelties in our choice of system. First, the S-matrix of the
undeformed theory is no longer just the identity matrix, since the seed theory is interacting. Second,
$ 4 $-particle scattering does not fix the structure of all
higher-point scattering since the seed theory is not integrable. In fact, as
one might suspect, for these cases renormalisability should run into
serious problems given the irrelevant nature of the deformation and
the absence of constraints imposed by quantum integrability on the
S-matrix. Nonetheless, we show by an explicit $ 1 $-loop computation
that due to low dimensionality, the $2 \rightarrow 2$ scattering
matrix at $ 1 $-loop uniquely fixes the renormalisation.

As observed by \citep{Rosenhaus:2019utc} and
\citep{Dey:2021jyl}, the classical deformed Lagrangian
(for massive free bosons or fermions) has only two mass scales,
viz.~mass $m$ and $\ttb$ coupling $\lambda$. However, the quantum
renormalised Lagrangian which reproduces the expected S-matrix turns
out to have multiple scales. Schematically, consider two interaction terms in the classical
deformed Lagrangian
 \begin{equation}
 	L_{\text{class.}} \subset \dots + \lambda \mc{O}_1 + \lambda \mc{O}_2 + \dots
 \end{equation}
In the renormalised theory, these two terms become
\begin{equation}
 	L_{\text{quant.}} \subset \dots + \alpha \mc{O}_1 + \beta \mc{O}_2 + \dots
 \end{equation}
 where 
\begin{equation}
	\begin{aligned} 
\alpha &= \lambda + \lambda ^2 f_1(\lambda,  m, \Lambda ) \ , \\
\beta  &= \lambda + \lambda ^2 f_2(\lambda , m, \Lambda ) \ ,
	\end{aligned} 
\end{equation}
with $\Lambda$ being the cut-off scale. This is certainly a novel
feature: the quantum theory sees the emergence of different couplings,
all of which flow to the same classical value, giving the impression
that the classical theory has a single coupling.

The mass term plays an important role here. Without the mass term, we
only see a renormalisation of the $\ttb$ coupling $ \lambda $, as expected in any
quantum field theory. The presence of the mass term allows for the
possibility of a new dimensionless parameter $m \lambda^2$ which is
crucial for the emergence of the new coupling. Similarly, if one has
additional scales in the undeformed action, like a coupling
$\lambda_{1}$ for a quartic interaction, then one can form
additional dimensionless parameters. New couplings emerging due to
quantum effects ought to reflect this fact. This is precisely what we
find by explicit computation.

The paper is organised as follows. We first discuss the (integrable)
case of chiral bosons in \pref{sec:chiral}. After this, we move on to
the (non-integrable) case of $\phi^{4}$ theory in \pref{sec:phi4}. We
conclude in \pref{sec:disc} with some comments and outline some
immediate open questions. A summary of various
notations and conventions used in this paper can be found in
\pref{app:conventions}. The Feynman rules and their derivation for the theory of chiral bosons in Sen's formalism are collected in \pref{app:feynmanrules}. Many of the computations in the text are
rather straightforward, but involve a number of integrals. To avoid
clutter in the main text, we have collected all of the integrals
used in \pref{app:integrals}.
 
\section{Renormalisation of Interacting Chiral Bosons} \label{sec:chiral}
\subsection{Free Chiral Bosons}
The action for a free chiral boson in Sen's formalism
\cite{Sen:2015nph,Sen:2019qit} is \footnote{Other known approaches to chiral bosons are \citep{Floreanini:1987as,Sonnenschein:1988ug,Pasti:1996vs}. For advantages of Sen's formalism over this other approaches we refer to the original papers \citep{Sen:2015nph,Sen:2019qit} (also see \citep{Andriolo:2020ykk,Chakrabarti:2020dhv}). Some recent works which have used Sen's formalism to great effect for self-dual field strengths in higher dimensions are \cite{Andriolo:2020ykk,Andriolo:2021gen,Gustavsson:2020ugb}.}
\begin{equation}
	S = \int \left[ \frac{1}{2} \dd\phi \wedge \star \dd \phi - 2
	A \wedge \dd \phi \right] \ .
\end{equation}
The field
$A_{\mu}$ is self-dual 1-form, i.e.~it satisfies the condition
$A^{\mu} = \epsilon^{\mu\nu}A_{\nu}$ and thereby has a single
independent component $A_{1}=-A_{0}=a$. The case of an anti-self-dual
$ 1 $-form is arrived at by making the replacement
$ A \rightarrow -B $ in the action, and anti-self-duality implies
$ B_{0} = B_{1} = b $. In the interest of brevity, we will only
discuss the the case of self-dual 1-forms, and the reader should note
that, \emph{mutatis mutandis}, what we say here goes through for the
case of anti-self-dual 1-forms as well.

It is natural to expect that chiral theories cannot support mass
terms. This expectation is correct: a term of the form $ A^{2} $ is
identically zero, since $ A^{2} = \epsilon^{\mu \nu}A_{\mu}A_{\nu}=0 $
by antisymmetry. The same, of course, is true of anti-self-dual gauge
fields.

The equations of motion are
\begin{equation}\label{key}
	\begin{aligned}
		\delta_{\phi} S = 0 \quad &\Rightarrow \quad \dd
		\left( \star \dd \phi + \dd \phi + 2 A \right) = 0 \ ,
		\\ \delta_{A} S = 0 \quad &\Rightarrow \quad \left(
		\star\dd \phi - \dd \phi \right) = 0 \ . \\
	\end{aligned}
\end{equation}
The free self-dual form $ (\star \dd \phi + \dd \phi + 2A) $ was shown
in \cite{Sen:2019qit} to decouple from the physical dynamics via a
Hamiltonian analysis of the constrained system. The physical field in
this model is $ A $, corresponding to a self-dual $ 1 $-form (i.e.~a
chiral boson) which satisfies a free equation of motion $ \dd A = 0 $.

Any classical solutions to this equation of motion can be parametrised
via a mode expansion
\begin{equation}\label{key}
	A_{\mu}(x) = \int \frac{\dd k}{2\pi} \frac{1}{2\omega} \left(
	\varepsilon_{\mu} \, \alpha(k) \, e^{ik\cdot x} + \text{c.c.}
	\right) \ ,
\end{equation}
for some choice of coefficients $ \alpha (k) $ provided that
\begin{equation}\label{key}
	k_{\mu} \varepsilon^{\mu} = 0 \ .
\end{equation}
This condition is a consequence of self-duality and the
equations of motion. The choice $ \varepsilon_{\mu} \propto k_{\mu} $
solves the above constraint, since $ k^{2} = 0 $ for massless
particles. The $ \varepsilon_{\mu} $ are similar to polarisation
vectors encountered when studying gauge fields, and this analogy is
occasionally profitable; for example, when considering scattering
processes, we must associate factors of $ \varepsilon_{\mu} $ to
external lines.

Integrating by parts and working in momentum space via the
two-dimensional Fourier transform, we have
\begin{equation}
	S = \frac{1}{2}\int \frac{\mathrm{d}^2 k}{(2\pi)^{2}} \,
	\left[\tilde{\phi}(k)\, k^{2} \, \tilde{\phi}(-k) - 4i
	\tilde{A}^{\mu}(k) \, k_{\mu} \, \tilde{\phi}(-k) \right]
	\ ,
\end{equation}
This is quadratic in the auxiliary field $ \tilde{\phi} $ and on completing
the squares (in a procedure that is fully analogous to
\cite[Sec.~7]{Sen:2015nph} which discusses the case of the self-dual
5-form in Type IIB supergravity) we can derive an expression for the
propagator of the self-dual gauge field. Define
\begin{equation}\label{key}
	\tilde{\varphi}(k) = \tilde{\phi}(k) - 2i \tilde{A}^{\mu}(k)
	\frac{k_{\mu}}{k^{2}} \ ,
\end{equation}
in terms of which the above action can be written as
\begin{equation}\label{key}
	S = \frac{1}{2} \int \frac{\dd^{2}k}{(2\pi)^{2}} \left[
	\tilde{\varphi}(k) \, k^{2} \, \tilde{\varphi}(-k) - 4
	\tilde{A}^{\mu}(k) \, \frac{k_{\mu} k^{\nu}}{k^{2}} \,
	\tilde{A}_{\nu}(-k) \right] \ .
\end{equation}
The term quadratic in $ \tilde{\varphi} $ decouples and we are left
with the action for physical fields
\begin{equation}\label{key}
	S = \int \frac{\dd^{2}k}{(2\pi)^{2}} \left[ \tilde{A}^{\mu}(k)
	\, \left(-2 \frac{k_{\mu} k^{\nu}}{k^{2}}\right) \,
	\tilde{A}_{\nu}(-k) \right] \ ,
\end{equation}
from which we can read off the free propagator of the
self-dual gauge field.  As pointed out in \cite{Sen:2015nph}, up to a
numerical factor this is precisely the kinetic term for self-dual $ 1
$-forms as discussed in \cite{Alvarez-Gaume:1983ihn}.

It is useful to keep in mind that the field $ A _{\mu } $ is self-dual
off-shell; to make
self-duality manifest, we rewrite the above expression as
\begin{equation}\label{key}
	S = \int \frac{\dd^{2}k}{(2\pi)^{2}} \left[ \left(
	\frac{\tilde{A}^{\mu}(k) + \epsilon^{\mu\rho}
		\tilde{A}_{\rho}(k)}{2} \right) \left(-2 \frac{k_{\mu}
		k_{\nu}}{k^{2}}\right) \left( \frac{\tilde{A}^{\nu}(-k) +
		\epsilon^{\nu\sigma} \tilde{A}_{\sigma}(-k)}{2}\right) \right] \ ,
\end{equation}
and on introducing the chiral projection operator
\begin{equation}\label{key}
	\mathcal{P}_{\pm}^{\mu\nu} = \frac{1}{2} \left( \eta^{\mu\nu}
	\pm \epsilon^{\mu\nu} \right) \ ,
\end{equation}
we can write the above expression as
\begin{equation}\label{key}
	S = \int \frac{\dd^{2}k}{(2\pi)^{2}} \frac{1}{2}  \left[
	\left(\mathcal{P}_{+}^{\mu\rho} \tilde{A}_{\rho}(k)\right)
	\left(-4 \frac{k_{\mu} k_{\nu}}{k^{2}}\right)
	\left(\mathcal{P}_{+}^{\nu\sigma}
	\tilde{A}_{\sigma}(-k)\right) \right] \ .
\end{equation}

The reader will note that this is the analogue of
\cite[eq.~(7.6)]{Sen:2015nph} for the case of self-dual $ 1 $-form
fields, in what is hopefully more transparent notation. We have also
included a factor of $ 1/2 $ in order to remain consistent with the
standard normalisations for field bilinears. We will see that the
presence of the projectors in the above expression is crucial to
ensure chiral propagation. This is easiest seen in lightcone
coordinates.

In terms of the self-dual $ 1 $-forms, the $ 2 $-point functions
relevant to this system are (as usual) given by $ i $ times the
inverse of the kinetic term. Componently, this means:
\begin{equation}\label{key}
		\left\langle \tilde{A}_{-}(k) \tilde{A}_{-}(-k)
		\right\rangle =\frac{-i}{4}
		\frac{k_{-}^{2}}{k^{2}} \ ,	
\end{equation}
and all other correlators identically vanish due to the projectors. A
similar expression is true for anti-self-dual gauge fields. Although
elementary, this brief exercise tells us that we really are working
with chiral fields.

\subsection{Relation to Free Scalars}
We have seen that Sen's formalism successfully reproduces the
$ 2 $-point functions of free chiral theories. It is useful, now, to
consider the question of interactions. In particular, we'd like to
explore the extent to which interacting chiral theories are related to
the more familiar subject of interacting non-chiral theories.

\subsubsection{Interacting Chiral Fields}

Consider a field of the form
\begin{equation}
	\label{eq:SDef}
	S _{\mu} = A _{\mu} + B _{\mu} \ ,
\end{equation}
where $ A _{\mu } $ and $ B _{\mu } $ are self- and anti-self-dual $ 1
$-form fields of the kind we encountered in the previous section.
Then, in terms of the projectors we introduced in the previous
section, we have
\begin{equation}
	\begin{aligned}
		\mathcal{P}^{\mu \nu}_{+} S _{\nu} = A _{\nu} \ , \\
		\mathcal{P}^{\mu \nu}_{-} S _{\nu} = B _{\nu} \ . \\
	\end{aligned}
\end{equation}
However, the field $ S _{\mu } $ itself has no definite chirality. Let us
further define the dual of $ S _{\mu} $ to be $ T ^{\mu} = \epsilon ^{\mu \nu} S _{\nu}
$; it is easy to see that
\begin{equation}
	T ^{\mu} = A ^{\mu} - B ^{\mu} \ ,
\end{equation}
on using (anti-)self-duality. The pair of fields $ (A,B) $ are chiral,
so we will refer to the pair of fields $ (S,T) $ as non-chiral.

Now, consider a theory of free left- and right-chiral fields. In
momentum space, their kinetic terms would ordinarily be written in
terms of the chiral pair as
\begin{equation}
	\int \frac{\mathrm{d}^{2}k}{(2 \pi)^{2}} \left[ \tilde{A} ^{\mu}(k)
	\left(-2 \frac{k _{\mu} k _{\nu}}{k ^{2}}\right) \tilde{A}
	^{\nu}(-k) + \tilde{B} ^{\mu}(k) \left(-2 \frac{k _{\mu} k
		_{\nu}}{k ^{2}}\right) \tilde{B} ^{\nu}(-k)\right] \ ,
\end{equation}
but in terms of the non-chiral pair, we can rewrite this as
\begin{equation}
	\int \frac{\mathrm{d}^{2}k}{(2 \pi)^{2}} \left[ \tilde{S} ^{\mu}(k)
	\left(- \frac{k _{\mu} k _{\nu}}{k ^{2}}\right) \tilde{S} ^{\nu}(-k) + \tilde{T}
	^{\mu}(k) \left(- \frac{k _{\mu} k _{\nu}}{k ^{2}}\right) \tilde{T}
	^{\nu}(-k) \right] \ .
\end{equation}

At this stage, all we have is a simple rewriting. Observe, however,
that in terms of the fields $ S $ and $ T $ any interaction that
couples left- and right-chiral fields can be written as
\begin{equation}
	A \cdot B = \eta ^{\mu \nu} A _{\mu} B _{\nu} = \frac{1}{2} S ^{2} = -
	\frac{1}{2} T ^{2} \ ,
\end{equation}
and more generally, for $ g \geq 2 $
\begin{equation}
	\label{eq:ABIntS}
	(A \cdot B ) ^{g} = \frac{1}{2 ^{g}} (S \cdot S) ^{g} \ .
\end{equation}
We could have chosen to write this in terms of the field $ T $ as well
--- this choice is equally acceptable. 
%The point to note is that when
%writing interactions in terms of the $ S $ and $ T $ fields, one of
%the fields can be thought of as free and decoupled. This decoupling is
%\emph{not} like the decoupling of the ghost field in Sen's formalism,
%and $ T $ will still contribute to the stress tensor, but in terms of
%the dynamics, it is inert.
We conclude from this discussion that at the level of the action, we
can write any theory of (equal numbers of) interacting left- and
right-chiral bosons as a theory of self-interacting scalars.

The relevance of this to our problem is as follows: from
\cref{eq:ABIntS} the leading interaction term introduced by the $
\ttb $ flow we studied in
\cite{Chakrabarti:2020dhv} can be written as
\begin{equation}
	\frac{\lambda}{4} (A \cdot B) ^{2} = \frac{\lambda}{16} (S \cdot S) ^{2} \ .
\end{equation}
Further, in order to match normalisations for the kinetic term we must
rescale $ S \rightarrow S/ \sqrt{2} $. So the final Lagrangian in
momentum space is
\begin{equation}
	\label{eq:SLagrangian}
	\int \frac{\mathrm{d}^{2}k}{(2 \pi)^{2}} \left[  \frac{1}{2} \tilde{S} ^{\mu}(k)
	\left(- \frac{k _{\mu} k _{\nu}}{k ^{2}}\right) \tilde{S} ^{\nu}(-k) +
	\frac{\lambda}{64} (\tilde{S}_\mu \tilde{S}^\mu)^{2} \right] \ .
\end{equation}
At this point, one can derive the Feynman rules for this theory and do
a standard textbook field theory computation to check agreement with
the CDD factor. It turns out, however, that there is a much more
direct way to establish a match with the CDD factor for this case,
which we will now describe.

\subsubsection{Ordinary Scalars Rewritten}

The action in \eqref{SLagrangian} is actually a familiar theory
written in an unfamiliar way, and in this section we make this
correspondence explicit. It is important to remember that by
definition in \eqref{SDef} the constituent field (say) $ A $ is more
akin to a ``field strength''. That is, in the more standard
formulations of chiral bosons, we would say that on-shell, $ A $ is a
conserved current. Of course, this perspective sits harmoniously with
Sen's formalism, since the equation of motion for $ A $ is
$ \mathrm{d} A = 0 $, which can be thought of as a conservation
equation.

The $\ttb$-deformed massless (ordinary) free boson (say, $\varphi$)
action is only a functional of the ``field-strength'' like variable
$V_\mu =\partial_\mu \varphi$. One can just as easily consider the
scattering of $V_\mu$ directly instead of considering the scattering
of $\varphi$. Of course, the Feynman rules need to be adjusted
accordingly, but the S-matrix, thanks to LSZ reduction, is impervious
to any field redefinition.

The Lagrangian for $\ttb$-deformed massless ordinary boson \cite{Cavaglia:2016oda} up to
first order in $\ttb$-coupling is
\begin{equation} 
\begin{aligned} \label{eq:Lagr_V}
	L &= -\frac{1}{2} \partial_\mu \varphi \partial^\mu \varphi -\frac{\lambda}{4} (\partial_\mu \varphi \partial^\mu \varphi)^2 \ , \\
	&= -\frac{1}{2} V_\mu V^\mu  -\frac{\lambda}{4} ( V_\mu V^\mu)^2 \ .
\end{aligned}
\end{equation} 
 The momentum space Feynman propagator for $V_\mu V_\nu$, as derived from the $\varphi \, \varphi$ propagator, is 
 \begin{equation}
			\langle V_{\mu}(-k) V_\nu(k) \rangle = -i\frac{k_\mu k_\nu}{k^2} \;.
 \end{equation}
This propagator is exactly the same as that of $\tilde{S}_\mu$ fields defined in previous section. Furthermore, the vertex factors as derived from \eqref{Lagr_V} matches exactly with the vertex factor derived from \eqref{SLagrangian}, provided we redefine the $\ttb$-coupling for chiral bosons as $\lambda \rightarrow \tilde{\lambda} = \lambda/16 $.

Therefore, to the leading order in $\ttb$-coupling, the deformed chiral bosons and deformed ordinary boson lead to a QFT with identical degrees of freedom and identical Feynman rules.\footnote{There is a redefinition of coupling constant, but since the coupling constant is dimensionful and there are no natural mass scales in the undeformed theory, this simply corresponds to a shift in choice of some reference scale, which suggests simply that there is no canonical choice of coupling. The physics is insensitive to this choice of scale.} Therefore, following the result already obtained for massless ordinary bosons in \citep{Rosenhaus:2019utc}, it automatically follows that up to first order, the $ 4 $-point S-matrix of the $\ttb$-deformed chiral boson theory agrees with the CDD factor. Of course, the CDD factor does not fix the real part of the $O(\lambda)$ contribution to S-matrix, which just like the case of free bosons, lead to an expected renormalisation of the $\ttb$ coupling.

At this point, it is perhaps tempting to conclude that the two theories will continue to be identical to all orders in the coupling. This, however, is not true. Lorentz invariance and the definition of the field variables guarantee that all interactions for chiral bosons will always be of the form $\sim (\tilde{S}_\mu \tilde{S}^\mu)^{n}$. However, the actual structure of the stress tensor for chiral bosons is very different from the stress tensor for free bosons. In fact, as explicitly established in \citep{Chakrabarti:2020dhv}, the closed form action is starkly different from that of the case of ordinary massless bosons. So at higher orders, the interaction terms will start to differ in two cases and one would need to resort to Feynman diagrammatics to compute the S-matrix.

Also, in \citep{Chakrabarti:2020dhv} it was shown that one can also find closed-form $\ttb$-deformed actions for arbitrary (but non-zero) number of left- and right-chiral bosons. In particular, for the case where there is a chiral asymmetry (i.e.~unequal numbers of left- and right-movers), one cannot re-write the action in terms of variables like $S_\mu$ and would have to work with the $A_\mu$ and the $B_\mu$ fields. We will not look at these cases in this paper, but for completeness and in the hope that they can be useful for other researchers looking to do perturbative computations in Sen's formalism, we give the Feynman rules in terms of $A$ and $B$ fields in \pref{app:feynmanrules}.

%What we have shown, therefore, is that we can begin with a theory of
%interacting left- and right-chiral fields and write it in terms of a
%theory of (derivatively) interacting scalars. 

%The crucial difference, however, is that in this reformulation, the
%$ 2 $-point functions of $ \eta $ are determined, fundamentally, from
%the $ 2 $-point functions of $ A $ and $ B $ that we derived earlier
%in this section. Neglecting interactions for the moment, it is easy to
%verify that the difference between this procedure and the procedure
%used in \cite{Rosenhaus:2019utc} when discussing massless scalars is
%precisely the case of ``mixed'' propagation: in their case, it is
%distributional and leads to local, divergent contributions to the
%scattering amplitudes and are consequently ignored; for us, it is
%identically zero. In either case, then, the results will match
%\emph{by construction}. 

%As long as we are doing perturbative computations, we will always
%reproduce the results of \cite{Rosenhaus:2019utc} since their Feynman
%rules can be mapped to ours. The only difference is in the definition
%of the $ \mathrm{T}\bar{\mathrm{T}} $ coupling constant. Still, in
%terms of this new coupling constant we are guaranteed to find
%scattering amplitudes that exponentiate to give the
%Castillejo-Dalitz-Dyson phase.

%Note, however, that our $ 4 $-particle interaction term is not
%integrable... \comment{This may be a good way to segue into a discussion
%	of non-integrable potentials and renormalisation.}
 The match with the CDD prediction vindicates the assumption made in \citep{Chakrabarti:2020dhv} regarding the viability of $\ttb$ deformation of a purely Lorentzian theory. In fact, this perhaps suggests that the definition of $\ttb$ deformation in terms of the exact quantum S-matrix, viz. \eqref{smatrix}, has wider applicability. The results of this section are perhaps unsurprising given that the undeformed theory was also an integrable quantum field theory without any mass scale. However, as pointed out in \citep{Rosenhaus:2019utc}, the CDD factor derivation holds for any undeformed theory, integrable or not. For non-integrable QFTs, the undeformed S-matrix is no longer just identity matrix and there will be particle production. However, that we are in low dimensions and that the deformed S-matrix is still given by a CDD phase multiplying the undeformed S-matrix together suggests that there should still be severe constraints on the renormalisation of such theories. This is what we turn to next, where we consider the theory of a massive scalar with quartic interaction as our undeformed seed theory.

\section{Renormalisation of $ \phi^{4} $ Theory} \label{sec:phi4}

Consider the Lagrangian for an interacting scalar field theory with a generic interaction:
\begin{align}
\mathcal{L}=\frac{1}{2}\left( \partial _\mu \phi\right) ^2 +V(\phi) \,.
\end{align}
The $\ttb$-deformed Lagrangian for this seed theory is \cite{Bonelli:2018kik}
\begin{align} \label{eq:ttbardeformed}
\tilde{\mathcal{L}}=-\frac{1}{2\lambda} \frac{1-2\lambda V}{1-\lambda V}+\frac{1}{2\lambda} \sqrt{\frac{(1-2\lambda V)^2}{(1-\lambda V)^2}+2\lambda \frac{\left( \partial _\mu \phi\right) ^2+2 V}{1-\lambda V}} \,,  
\end{align}
where $\lambda$ is the $\ttb$ coupling.  In this section,  we consider massive scalar theory with a bare $\phi ^4$ potential: $V=\frac{1}{2}m^2 \phi ^2 +\lambda _1 \phi ^4$.  Here we assume that the mass $m$ is physical and therefore does not require any renormalisation.
The relevant Lagrangian for $2 \rightarrow 2$ scattering can be obtained by expanding \eqref{ttbardeformed} upto quadratic order in both of the couplings
\begin{align}
\mathcal{\tilde{L}} =\frac{1}{2}\left( \partial _\mu \phi\right) ^2 &+\frac{m^2 \phi ^2}{2} +\lambda _1 \phi ^4+\frac{1}{4} \lambda  m^4 \phi ^4-\frac{\lambda  }{4}\left( \partial _\mu \phi \partial ^\mu \phi\right) ^2 \cr
&+\lambda \lambda _1 m^2 \phi ^6 +\lambda ^2 \left( \frac{1}{4}\left( \partial _\mu \phi \partial ^\mu \phi \right) ^3+\frac{m^2}{8}\left( \partial _\mu \phi \partial ^\mu \phi\right) ^2 \phi ^2+\frac{m^6}{8}\phi ^6\right)\,.  \label{eq:barelagrangian}
\end{align}
It is useful to note that the mass dimensions of the $\ttb$ and bare $\phi ^4$ couplings are $-2$ and $2$ respectively.

In order to obtain the Feynman rules for the $\ttb$-deformed theory, we have to rewrite the Lagrangian in \eqref{barelagrangian} in Lorentzian signature.  This can be done by substituting $(m^2, \lambda _1,\lambda) \rightarrow (-m^2,-\lambda _1,-\lambda)$ and we obtain
\begin{align}
\tilde{\mathcal{L}}_{\text{L}}=-\frac{1}{2}\left( \partial _\mu \phi\right) ^2-&\frac{m^2 \phi ^2}{2} -\lambda _1 \phi ^4-\frac{1}{4} \lambda  m^4 \phi ^4+\frac{\lambda  }{4}\left( \partial _\mu \phi \partial ^\mu \phi\right) ^2 \cr
&-\lambda \lambda _1 m^2 \phi ^6 +\lambda ^2 \left( \frac{1}{4}\left( \partial _\mu \phi \partial ^\mu \phi \right) ^3-\frac{m^2}{8}\left( \partial _\mu \phi \partial ^\mu \phi\right) ^2 \phi ^2-\frac{m^6}{8}\phi ^6\right)\,.
\end{align}
From this Lagrangian,  the Feynman rules can be read off:
\begin{align}
\text{propagator}: & \quad \frac{-i}{p^2-m^2}\,,\\
\text{4-scalar vertex}: &\quad  -i\Big(\lambda _1+\frac{\lambda m^4}{4}\Big) +\frac{i\lambda}{4\cdot 3} \Big[ (p_1 \cdot p_2)(p_3\cdot p_4)+(2 \leftrightarrow 3)+(2 \leftrightarrow 4)\Big] \,. \label{eq:4scalar}
\end{align}
Parametrizing the components of external momenta in terms of individual rapidity $\theta _i$ 
\begin{align}
p^0_i=m\cosh \theta _i \,, \quad p^1_i=-m\sinh \theta _i \,, \label{eq:thetaparametrization}
\end{align}
we express the Mandelstam variables, in $(-,+)$ signature for all ingoing external momenta as follows
\begin{align}
s=(p_1+p_2)^2=2m^2(1-\cosh \theta _{12})\,, \\ u=(p_1+p_3)^2=2m^2(1-\cosh \theta _{13})\,,\\
t=(p_1+p_4)^2=2m^2(1-\cosh \theta _{14})\,.\label{eq:mandelstam}
\end{align}
Here we define relative rapidities as $\theta _{ij}:=\theta _i-\theta _j$.  The parametrization \eqref{thetaparametrization} also implies that various momentum scalar products can be written as
\begin{align}
 p_1\cdot p_2 =-m^2 \cosh \theta _{12} \,, \quad p_1\cdot p_4=-m^2\cosh \theta _{14} \,, \quad p_1\cdot p_3 =-m^2 \cosh \theta _{13} \,.
\end{align}
All other momentum scalar products can be expressed in terms of the above scalar products. 
\subsection{$2\rightarrow 2$ scattering}
In the case of four particle scattering in $(1+1)$ dimensions,  one can always set the spatial component of the momentum vector as
\begin{align}
p_1=-p_3\,,\quad p_2=-p_4\,,
\end{align}
leading to the temporal component ($p^0\equiv E$) to be related as $E_1=\pm E_3$ and $E_2=\pm E_4$.  By setting $E_1=E_3$ and using $E_1^2=m^2+p_1^2$,  any one of the three Mandelstam variables (say $u$) can be set to zero
\begin{align}
u=(p_1+p_3)^2=2m^2+2(-E_1E_3+p_1p_3)=0\,.
\end{align}
In the following analysis,  we set $u=0$ (or $\theta _{13}=0$).  Then the onshell condition $s+t=4m^2$ for $u=0$,  relates the two other relative rapidities as  $\theta _{14}=(i\pi -\theta _{12})$. Therefore the Mandelstam variables become function of a single relative rapidity factor:
\begin{align}
s=2m^2(1-\cosh \theta)\,, \quad u=0\,, \quad t=2m^2(1+\cosh \theta)\,,
\end{align}
where we rewrite $\theta _{12} \equiv \theta $ to avoid clutter.
At tree level the 4-scalar vertex for the integrable theory can now can be expressed as
\begin{align}
V_{\phi \phi \phi \phi}= -i\left(\lambda _1+\frac{\lambda m^4}{4}\right) +\frac{i\lambda m^4}{12} \Big[2 \cosh ^2 \theta +1 \Big] \,.  
\end{align}

 \subsection{Tree amplitude}
\begin{figure} 
\begin{center}
\includegraphics[scale=.7]{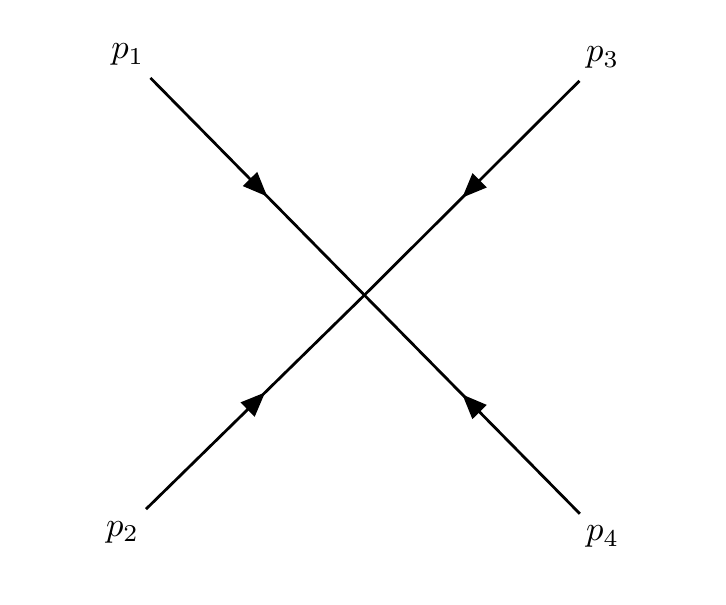}
\end{center}
\caption{Contact scattering diagram for massive $\phi ^4$ theory at tree level} \label{fig:tree}
\end{figure}

The tree level amplitude (\pref{fig:tree}) now can be expressed as function of rapidity
\begin{align} \label{eq:4-pt-tree}
\mathcal{A}^{\text{tree}}_4(\theta) &=-i4! \left[ \left( \lambda _1+\frac{\lambda m^4}{4}\right)- \frac{\lambda}{12} m^4 (2\cosh ^2\theta +1) \right]\nonumber\\
&=-24 i \lambda _1 +4i \lambda m^4 \sinh ^2\theta \,.  
\end{align}
This matches with result of \citep{Rosenhaus:2019utc} by using a simple substitution:
\begin{align}
\frac{1}{4}\lambda m^4 \longrightarrow \lambda _1+\frac{1}{4}\lambda m^4\,.
\end{align}

\subsection{From the CDD factor} \label{sec:cddanalysis}
We can quickly check that the tree level amplitude computed in the previous section is precisely what one would expect based on the CDD factor. Recall, The quantum S-matrix is related to the scattering amplitude via the following relation \citep{Rosenhaus:2019utc,Dey:2021jyl}
\begin{align}
S(\theta )=1+\frac{T}{4m^2\sinh \theta }\,.
\label{eq:smatrix-amplitude}
\end{align}
At linear order in $\ttb$ coupling this gives
\begin{align}
S_{\lambda}= S_{0} \left(1+i\lambda m^2 \sinh \theta \right)\,.
\end{align}
Therefore the amplitude for the $\ttb$-deformed theory $ T_{\lambda} $ is related to the amplitude of the undeformed seed theory $T_{0}$ as
\begin{align}
T_{\lambda}=T_{0}+i\lambda m^2 T_{0}  \sinh \theta +4i\lambda m^4 \sinh ^2 \theta  \,.
\end{align} 
If the bare $\phi ^4$ coupling is set to zero then $T_{0}=0$, which corresponds to a seed theory describing free propagation and we recover the tree amplitude computed in \citep{Rosenhaus:2019utc} for the $\ttb$-deformed theory.  With bare $\phi ^4$ coupling turned on, the amplitude for the undeformed seed theory is  $T_{0}=-24i\lambda _1$.  In this case, the tree amplitude for the $\ttb$-deformed theory is
\begin{align}
\mathcal{A}^{\text{tree}}_4(\theta) \equiv T_{\lambda}=-24i\lambda _1+4i\lambda m^4 \sinh ^2 \theta    \,,
\end{align}
which precisely matches the result obtained by explicit computation in \eqref{4-pt-tree}.

\subsection{1-loop amplitude} \label{sec:1loopphi4}
\begin{figure}[h!]
\begin{center}
\includegraphics[scale=.9]{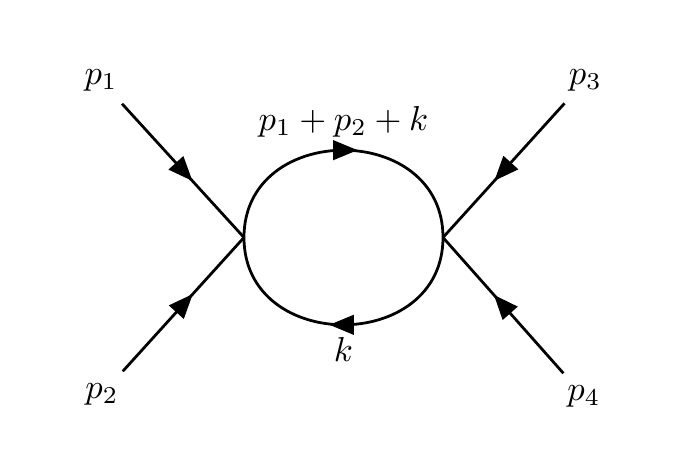}
\end{center}
\caption{1-loop bubble diagram for massive $\phi ^4$ theory.}
\label{fig:fig1}
\end{figure}

\begin{figure}
\begin{center}
\includegraphics[scale=.9]{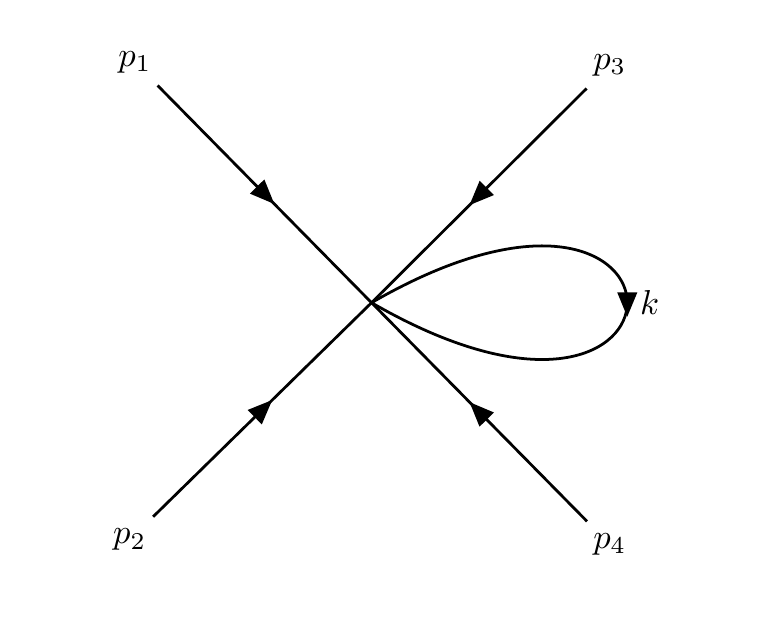}
\end{center}
\caption{1-loop tadpole diagram for massive $\phi ^4$ theory.}
\label{fig:fig2}
\end{figure}

Since the bare Lagrangian in \eqref{barelagrangian} has sextic terms proportional to $\lambda ^2$ and $\lambda \lambda _1$,  there exists two classes of diagrams that contribute to the second order S-matrix for  $2\rightarrow 2$ scattering: first,  the 1-loop bubble diagram given in \pref{fig:fig1} and second, we have to evaluate the tadpole diagram in \pref{fig:fig2}.

We begin with the 1-loop bubble diagram in $s$-channel (see \pref{fig:fig1}).  This is given  by the following integral
\begin{align}
\mathcal{A}^{1\text{-loop}}_{s}[1,2,3,4]=\int \frac{\mathrm{d}^2k}{(2\pi)^2} \frac{N_{s}}{(k^2-m^2)[(k+p_1+p_2)^2-m^2]} \,,
\end{align}
where the numerator is 
\begin{align}
N_{s} =\left[i \chi +\frac{i\lambda}{12} \Big\lbrace (p_1\cdot p_2)(k \cdot \overline{k+P}) +(p_1\cdot k)(p_2\cdot \overline{k+P})+(p_1\cdot \overline{k+P})(p_2\cdot k)  \Big\rbrace \right]\cr \times \left[i \chi +\frac{i\lambda}{12} \Big\lbrace (p_3\cdot p_4)(k \cdot \overline{k+P}) +(p_3\cdot k)(p_4\cdot \overline{k+P})+(p_3\cdot \overline{k+P})(p_4\cdot k)  \Big\rbrace \right]\,. \label{eq:phi4integrand}
\end{align}
Here $\chi :=\left( \lambda _1+\frac{\lambda m^4}{4}\right)$ and $P:=(p_1+p_2)$.  Note that the derivative terms are adding up as opposed to tree amplitude (where all momenta are ingoing) because one of the two loop momenta will always be outgoing at the vertex. 

We rewrite the integrand by classifying the contributions from diagrams involving $ (a) $ no derivative, $ (b) $ single derivative, and $ (c) $ both derivative interactions at the vertices
\begin{align}
N_s=(i \chi)^2 +(i \chi)\left(\frac{i\lambda}{12}\right) \mathcal{M}^{(b)} +\left(\frac{i\lambda}{12}\right)^2 \mathcal{M}^{(c)} \,.
\end{align}

The $ s $-channel contribution from diagrams involving no derivatives is
\begin{align}
\mathcal{A}^{(a)}_s =- 72 \left( \lambda _1+\frac{\lambda m^4}{4}\right)^2 \frac{( \pi +i\theta )}{ \pi m^2 \sinh \theta _{12}} \,.\label{eq:noderivative}
\end{align}  
The contribution from the $u$- and $t$-channel can be found by simply setting $\theta =0$ and $\theta =(i \pi -\theta )$ respectively.  We combine those results to obtain the total contribution without derivative interactions at the vertices:
\begin{align}
\mathcal{A}^{(a)}=\frac{72}{ \pi  m^2} \left(\lambda _1 +\frac{\lambda  m^4}{4}\right)^2 \left(i- \frac{\pi}{\sinh \theta }\right) \,.
\end{align}
Similarly, the contribution from diagrams with single derivative interaction is
\begin{align}
\mathcal{A}^{(b)} &=-\frac{12i m^2}{\pi}\lambda \left( \lambda _1+\frac{\lambda m^4}{4}\right)  \left[2 \log \left(\frac{m^2}{\Lambda ^2}\right)+i \pi  \left(2 \sinh \theta +\frac{3}{\sinh \theta}\right)\right]\,,\label{eq:singlederivative}
\end{align}
and diagrams with two derivative interactions is
\begin{equation}
	\begin{aligned}
		\mathcal{A}^{(c)}=\frac{i\lambda ^2 m^6}{12\pi}&\left[i\pi \frac{6(2+\cosh (2\theta))^2}{\sinh \theta} +8+57\frac{\Lambda ^2}{m^2}+90\log \left(\frac{m^2}{\Lambda ^2}\right) \right. \\
		&\qquad \qquad \qquad \qquad \left. +8\cosh (2\theta)\left(-2+3\frac{\Lambda ^2}{m^2}+3\log \frac{m^2}{\Lambda ^2}\right)  \right]\,.\label{eq:doublederivative}
	\end{aligned}
\end{equation}
Adding all three contributions, we obtain the amplitude from the $ 1 $-loop bubble diagram as
\begin{equation}
	\begin{aligned}
		\mathcal{A}_{\text{bubble}}=\frac{1}{12 \pi  m^2}&\left[i \lambda ^2 m^6 \left\lbrace 57 \Lambda ^2+8 \cosh (2 \theta ) \big(3 \Lambda ^2+3 m^2 \log \left(\frac{m^2}{\Lambda ^2}\right)-2 m^2\big )+62 m^2 \right. \right. \\
		&\qquad \qquad \left. +18m^2 \log \left(\frac{m^2}{\Lambda ^2}\right)+24 i \pi m^2 \sinh ^3\theta \right\rbrace \\
		&+144 \lambda _1 \Big\lbrace \lambda _1 (6 i-6 \pi  \text{csch}\theta )\\
		&\qquad \qquad \left. +\lambda  m^4 \big(2 \pi  \sinh \theta -2 i \log \left(\frac{m^2}{\Lambda ^2}\right)+3 i\big)\Big\rbrace \right]\,.
	\end{aligned}
\end{equation}
This matches with the result given in \citep{Rosenhaus:2019utc} for 1-loop bubble diagram when we set $\lambda _1 =0$.

\subsection{Tadpole contribution}
 The relevant terms in the $\ttb$-deformed Lagrangian (in Lorentzian signature) for massive $ \phi ^4$ theory that give rise to tadpoles are
\begin{align}
\tilde{\mathcal{L}}_{\lambda ^2,\lambda \lambda _1} =-\lambda \lambda _1 m^2 \phi ^6+\lambda ^2 \left( \frac{1}{4}\left( \partial _\mu \phi \partial ^\mu \phi\right) ^3-\frac{m^2}{8}\left( \partial _\mu \phi \partial ^\mu \phi\right) ^2 \phi ^2-\frac{m^6}{8}\phi ^6\right)\,.
\end{align}
In order to find the tadpole contribution, we first write down the effective Lagrangian where we Wick contract any two of the external scalars.  During the Wick contraction, it is important to keep track of how many ways one can perform this operation.  For terms with $\phi ^n$ polynomial structure,  this number is simply ${}^nC_2$.  For terms with derivative, these numbers are the following
\begin{align}
\left(\partial _\mu \phi \partial ^\mu \phi \right)^3 \longrightarrow 9 \,, \quad  \left(\partial _\mu \phi \partial ^\mu \phi \right)^2 \longrightarrow 4 \,.
\end{align}
After Wick contraction, we obtain the following Lagrangian for the  tadpole diagram 
\begin{equation}
	\begin{aligned}
		\tilde{\mathcal{L}}_{\text{tadpole}}&=\left(\tfrac{9}{4} \lambda ^2\langle  \partial _\mu \phi \partial ^\mu \phi \rangle -\tfrac{\lambda ^2 m^2}{8}\langle \phi ^2\rangle\right) \left( \partial _\mu \phi \partial ^\mu \phi \right)^2 -\tfrac{\lambda ^2 m^2}{2} \phi ^2 \left( \partial _\mu \phi \partial ^\mu \phi \right) \langle  \partial _\mu \phi \partial ^\mu \phi \rangle \\
		&\quad -15\left( \lambda \lambda _1 m^2+\tfrac{\lambda ^2 m^6}{8} \right)\phi ^4  \langle \phi ^2\rangle \,.
	\end{aligned}
\end{equation}
The Feynman rule for the vertex can be easily derived from the above Lagrangian 
\begin{align}
V_{\text{tadpole}}= \frac{i\kappa _1}{ 3} \Big[ (p_1 \cdot p_2)(p_3\cdot p_4) &+(2\leftrightarrow 3) +(2\leftrightarrow 4)\Big]\cr
&+i\kappa _2 \left\lbrace (1;234)+(2;34)+(p_3\cdot p_4)\right\rbrace -i\kappa _3 \,,
\label{eq:tadpolefeynman}
\end{align}
where the coefficients $\kappa _i$'s are 
\begin{align}
\kappa _1 &:= \left(\tfrac{9}{4} \lambda ^2\langle  \partial _\mu \phi \partial ^\mu \phi \rangle -\tfrac{\lambda ^2 m^2}{8}\langle \phi ^2\rangle\right) \,,\\
\kappa _2 &:=\tfrac{\lambda ^2 m^2}{2} \langle  \partial _\mu \phi \partial ^\mu \phi \rangle \,,\\
\kappa _3&:=15\left( \lambda \lambda _1 m^2+\tfrac{\lambda ^2 m^6}{8} \right)\langle \phi ^2\rangle \,,
\end{align}
and additionally we have used the following notations to denote various momentum scalar products appearing in the vertex
\begin{align}
(1;234)&:=(p_1\cdot p_2)+(p_1\cdot p_3)+(p_1\cdot p_4)\,,\\
(2;34)&:=(p_2\cdot p_3)(p_2\cdot p_4)\,.
\end{align}
We set $u=0$ for $2\rightarrow 2$ scattering in two dimensions and rewrite the Feynman rule in terms of relative rapidity $\theta$
\begin{align}
V_{\text{tadpole}}= \frac{i\kappa _1}{ 3} m^4(2\cosh ^2\theta +1)-2im^2\kappa _2  -i\kappa _3 \,.
\end{align}
Making use of this Feynman rule, we obtain the contribution to the amplitude from the tadpole diagram as 
\begin{equation}
	\begin{aligned}
		\mathcal{A}_{\text{tadpole}}=4!i\lambda ^2 m^4 &\left[\frac{(1+2\cosh
			^2\theta)}{3}\left(\tfrac{9}{4} \langle  \partial _\mu \phi \partial ^\mu \phi \rangle -\tfrac{ m^2}{8}\langle \phi ^2\rangle\right) \right. \\
		&\qquad \qquad-\tfrac{1}{6} \langle  \partial _\mu \phi \partial ^\mu \phi \rangle
		-15\big(\tfrac{\lambda _1}{\lambda m^2}+\tfrac{m^2}{8}\big)\langle \phi ^2\rangle \bigg]\,.
	\end{aligned}
\end{equation}
Note that, the combinatoric factor for the diagrams with contact interaction $\phi ^2 \partial _\mu \phi \partial ^\mu \phi $ is 4 as opposed to $4!$ of the other two classes of interactions.  Using the expressions for the following two point correlators
\begin{align}
\langle \phi ^2\rangle =-\frac{1}{4\pi} \log \left(\frac{m^2}{\Lambda ^2}\right)\,, \quad  \langle  \partial _\mu \phi \partial ^\mu \phi \rangle=-\frac{1}{4\pi}\left[m^2 \log \left(\frac{m^2}{\Lambda ^2}\right)+\Lambda ^2\right]\,,
\end{align}
we obtain the tadpole amplitude as 
\begin{equation}
	\begin{aligned}
		\mathcal{A}_{\text{tadpole}}&=\frac{ i \lambda ^2 m^6}{4 \pi }\big[15-17 \cosh (2 \theta  )\big]\log \left(\frac{m^2}{\Lambda ^2}\right)\\
		&\qquad -\frac{i \lambda ^2 \Lambda ^2 m^4}{2 \pi }\big[9 \cosh (2 \theta  )+16\big]
		+\frac{90 i \lambda  \lambda _1 m^2}{\pi } \log \left(\frac{m^2}{\Lambda ^2}\right)\,. \label{eq:tadpole}
	\end{aligned}
\end{equation}

\subsection{renormalisation }
The amplitude at $\mathcal{O}(\lambda ^2, \lambda \lambda _1,\lambda _1^2)$ is a sum of the 1-loop bubble and the tadpole contributions: 
\begin{equation}
	\begin{aligned}
		\mathcal{A}^{(2)}=\frac{1}{12 \pi} &\left[-im^6\lambda ^2 \left\lbrace 39 \frac{\Lambda ^2}{m^2}+\cosh 2 \theta \left(30 \frac{\Lambda ^2}{m^2}+27 \log \left(\frac{m^2}{\Lambda ^2}\right)+16\right) \right. \right. \\
		&\qquad \qquad\qquad  \left. {\color{black}-24 i \pi \sinh ^3\theta}
		-63  \log \left(\frac{m^2}{\Lambda ^2}\right)-62   \right\rbrace \\
		&\qquad +72 \lambda  \lambda _1  \left\lbrace  {\color{black}4\pi m^2  \sinh \theta }+11i m^2  \log \left(\frac{m^2}{\Lambda ^2}\right)+i6m^2 \right\rbrace \\
		&\qquad \qquad \qquad \qquad \qquad \left. -{\color{black}864 \frac{\lambda _1^2}{m^2} (\pi  \text{csch}\theta -i)}\right]\,.  \label{eq:1-loopresult}
	\end{aligned}
\end{equation}

The S-matrices of the $\ttb$-deformed and undeformed seed theory  are related to each other by the following relation upto quadratic order in $\ttb$ coupling $\lambda$
\begin{align}
S_{\lambda}= S_{0} \left(1+i\lambda m^2 \sinh \theta -\frac{1}{2}\lambda ^2m^4 \sinh ^2 \theta \right)\,.
 \label{eq:cddfactor}
\end{align}
Repeating the analysis of \Secref{cddanalysis} and using \eqref{smatrix-amplitude}, we find the relation between the scattering amplitudes of the two theories
\begin{align}
T_{\lambda}=T_{0} +4i\lambda m^4\sinh ^2\theta -2\lambda ^2 m^6\sinh ^3 \theta +iT_{0} \lambda m^2\sinh \theta +\mathcal{O}(\lambda ^2 \lambda _1,\ldots)\,.
\end{align}
The amplitude of the undeformed massive $\phi ^4$ theory upto quadratic order in bare $\phi ^4$ coupling is
\begin{align}
T_{0}=-24i\lambda _1+\frac{72\lambda _1^2 }{\pi m^2} (i-\pi \text{csch} \theta)\,.
\end{align}
Therefore,  the amplitude of the $\ttb$-deformed theory at $\mathcal{O}(\lambda ^2, \lambda \lambda _1, \lambda _1^2)$ is given by
\begin{align}
T_{\lambda}^{(2)}=-2\lambda ^2 m^6\sinh ^3 \theta +24 \lambda \lambda _1 m^2\sinh \theta +\frac{72\lambda _1^2 }{\pi m^2} (i-\pi \text{csch} \theta)\,.\label{eq:2ndordercddfactor}
\end{align}
Now let us first consider the terms with $\ttb$ coupling $\lambda$ which are purely real and exactly matches with the real part of the  amplitude in \eqref{1-loopresult}. However, the imaginary part of the amplitude can not be fixed by \eqref{2ndordercddfactor} and should cancel on adding counterterms to the bare Lagrangian.  Let us collect all the imaginary pieces in \eqref{1-loopresult}
\begin{equation}
	\begin{aligned}
		\text{Im}\,\mathcal{A}^{(2)}&=\frac{-i}{12 \pi} \left[m^6\lambda ^2 \left\lbrace 69 \frac{\Lambda ^2}{m^2}-36  \log \left(\frac{m^2}{\Lambda ^2}\right)-46 \right. \right. \\
		&\qquad \qquad \qquad \qquad \left. +\sinh ^2 \theta \left(60 \frac{\Lambda ^2}{m^2}+54 \log \left(\frac{m^2}{\Lambda ^2}\right)+32\right) \right\rbrace \\
		&\qquad \qquad \qquad \qquad \qquad \qquad \qquad \left. -72 \lambda  \lambda _1  m^2\left\lbrace   11 \log \left(\frac{m^2}{\Lambda ^2}\right)+6 \right\rbrace  \right]\,. 
	\end{aligned}
\end{equation}
We assume that the both the bare $\phi ^4$ and $\ttb$ coupling are classical couplings and that both of them get renormalised. Therefore we add counter terms at $\mathcal{O}(\lambda ^2, \lambda \lambda _1)$ to the bare Lagrangian in \eqref{barelagrangian}, we have
\begin{align}
\mathcal{\tilde{L}}_{\text{renorm.}} =\frac{1}{2}\left( \partial _\mu \phi\right) ^2&+\frac{m^2 \phi ^2}{2} +\left(\lambda _1 +\frac{\lambda  m^4}{4} \right) \phi ^4-\left(\frac{\lambda}{4} -\gamma \lambda ^2\right) \left( \partial _\mu \phi \partial ^\mu \phi \right)^2 +(\alpha \lambda ^2 +\beta \lambda \lambda _1 ) \phi ^4  \cr
&+\lambda \lambda _1 m^2 \phi ^6 +\lambda ^2 \left( \frac{1}{4}\left( \partial _\mu \phi \partial ^\mu \phi \right)^3+\frac{m^2}{8}\left( \partial _\mu \phi \partial ^\mu \phi \right)^2 \phi ^2+\frac{m^6}{8}\phi ^6\right)\,.  \label{eq:ctlagrangian}
\end{align}
We do not consider any counterterm at $\mathcal{O}(\lambda _1^2)$ as the this is completely fixed by the last term of \eqref{2ndordercddfactor}, since the CDD factor does not involve bare $\phi ^4$ coupling.
We only need to consider the tree level contribution to amplitude due to these counter terms as they are already at $\mathcal{O}(\lambda ^2,\lambda \lambda _1)$. The amplitude due to counterterms is 
\begin{align}
\mathcal{A}_{\text{count.}}&=24i \left[ -(\alpha \lambda ^2+\beta \lambda \lambda _1)-\frac{ \gamma}{3}\lambda ^2 m^4 (2\sinh ^2\theta +3) \right]\,,\\
&=-24i\left[( \alpha +\gamma m^4)\lambda ^2 +\beta \lambda \lambda _1+\frac{2}{3} \gamma \lambda ^2m^4 \sinh ^2\theta \right]\,.
\end{align}
The additional sign is due to change in metric signature.  In order to get rid of the imaginary part of the amplitude, we set
\begin{align}
\gamma &=-\frac{m^2}{32 \pi} \left[10 \frac{\Lambda ^2}{m^2}+9 \log\left(\frac{m^2}{\Lambda ^2}\right)+\frac{16}{3}\right]\,,\\
\beta &=\frac{m^2}{4\pi} \left[11\log \left(\frac{m^2}{\Lambda ^2}\right)+6\right]\,,
\end{align}
and $\alpha$ gets automatically fixed to be
\begin{align}
\alpha =\frac{m^6}{96\pi} \left[\frac{94}{3}+7\frac{\Lambda ^2}{m^2}+39\log\left(\frac{m^2}{\Lambda ^2}\right)\right]\,.
\end{align}
This analysis shows that the renormalised $\ttb$ Lagrangian involves two ``new'' couplings associated to $\phi ^4$ and $\left( \partial _\mu \phi \partial ^\mu \phi\right) ^2$ interactions:
\begin{align}
\chi _{\text{renorm.}}&=\chi _{\text{bare}} +\sum _{n \geq 2} a_n \lambda ^n +\sum _{n \geq 1} b_n (\lambda \lambda _1)^n\,,\\
\omega _{\text{renorm.}} &= \sum _{n \geq 1} c_n \lambda ^n\,.
\end{align}
Here $\chi _{\text{bare}}=\left(\frac{\lambda _1}{m^4}+\frac{\lambda }{4}\right)$ and $c_1=1$.  The renormalised Lagrangian can be recast as 
\begin{align}
\mathcal{\tilde{L}}_{\text{renorm.}} =\frac{1}{2}\left( \partial _\mu \phi\right) ^2+\frac{1}{2}m^2 \phi ^2 +\frac{1}{4}\chi _{\text{renorm.}}m^4 \phi ^4-\frac{1}{4}\omega _{\text{renorm.}} \left(\partial _\mu \phi \right) ^2 +\cdots\,,
\end{align} 
where at 1-loop,  the $2\rightarrow 2$ scattering matrix fully fixes the coefficients in the series expansion of renormalised couplings as
\begin{align}
a_2&=\frac{m^2}{24\pi} \left[\frac{94}{3}+7\frac{\Lambda ^2}{m^2}+39\log\left(\frac{m^2}{\Lambda ^2}\right)\right]\,,\\
b_1 &=\frac{1}{\pi m^2} \left[11\log \left(\frac{m^2}{\Lambda ^2}\right)+6\right]\,,\\
c_2&=\frac{m^2}{8 \pi} \left[10 \frac{\Lambda ^2}{m^2}+9 \log\left(\frac{m^2}{\Lambda ^2}\right)+\frac{16}{3}\right]\,.
\end{align}
The results in this section show that the bare $\phi ^4$ interaction introduces a new family of dimensionless couplings $(\lambda \lambda _1)^n$, which starts appearing from the first order in perturbation theory with coefficients $b_n$, comprised within $\chi _{\text{renorm}}$.  This feature supports our claim in the introduction that additional scales (here $\lambda _1$) in the seed action can be used to form new dimensionless parameters along with the $\ttb$ parameter $\lambda$, which then contributes to the emergence of novel couplings in the quantum mechanically corrected theory.

A rudimentary, but non-trivial consistency check of our calculation is we should reproduce the answer obtained in \cite{Rosenhaus:2019utc} if we set the bare quartic coupling $\lambda_1=0$. This can be checked rather straightforwardly.\footnote{Since \cite{Rosenhaus:2019utc} gave their result without the finite pieces, one can only confirm agreement with their calculations without the finite pieces.} In our computation we have kept all the finite pieces arising in our renormalisation scheme. As indicated in \cite{Rosenhaus:2019utc}, the knowledge of the finite pieces are important in the calculation of correlation functions with renormalised Lagrangian. While the expectations from dimension counting are vindicated by this explicit computation, we note that the situation becomes more interesting for higher-point scattering amplitudes. A systematic approach to analyse the higher point scattering will most definitely shed light on this intriguing interplay of renormalisation and the knowledge of exact quantum S-matrix.
\section{Discussion} \label{sec:disc}

The central theme of this paper was the study of constraints imposed on the renormalisation of $\ttb$-deformed theories coming from knowledge of the exact deformed S-matrix. We extended previous investigations into this theme in two directions. Both these directions involved the study of the relationship between deformed and undeformed S-matrices.

The first deformed theory we looked at was an integrable field theory describing a pair of (left- and right-)chiral bosons. The deformation of this theory shared similar qualitative features as that of an ordinary massless (free) boson \citep{Rosenhaus:2019utc}. The strong constraints of integrability along with a lack of any mass scale in the bare Lagrangian inevitably leads to only an expected renormalisation of the $\ttb$ coupling $\lambda$. In this case too, the S-matrix takes exactly the form dictated by the CDD factor. This computation contains some novelties, in that the seed theory has some unusual features. While it has been very clear from the outset that there is a meaningful way to extract Feynman rules for these kinds of theories (viz.~self-dual $ p $-forms), and is already implicit in \cite{Sen:2015nph}, to the best of our knowledge this is the first time an explicit derivation of Feynman rules are presented starting from an action. We expect these technical results to be useful for studies of (anti)-self-dual fields and their interactions in higher dimensions as well.

The second example we tackled involved an exploration of the constraints imposed by the CDD factor on renormalisability of a non-integrable quantum field theory. The undeformed theory we considered was a massive $\phi^4$ theory, and therefore contained \emph{two} different mass scales in the bare Lagrangian.\footnote{Recall that in two dimensions the $\phi^4$ coupling is dimensionful.} We studied the $ 4 $-point scattering up to $ 1 $-loop, for which the computation was somewhat cumbersome. Nevertheless, the end result of the computation was worthwhile: we found that the $ 1 $-loop S-matrix for $ 2 \rightarrow 2 $ scattering matches precisely with CDD factor. The quantum action has new couplings that flow to the $\ttb$ coupling at tree level, echoing similar observations made for integrable field theories in \citep{Rosenhaus:2019utc,Dey:2021jyl}. In fact, we find an additional contribution to one of the emergent couplings (compared to the case of the free massive boson) which can be traced to the fact that there is an additional mass scale in the undeformed theory, viz.~the $\phi^4$ coupling. This is consistent with the intuition that the existence of mass scales in the undeformed theory is crucial for the emergence of new couplings in the quantum action.

There are several questions that follow naturally from the results of this paper. Firstly, unlike the integrable quantum field theories, non-integrable QFTs do not have their $ n $-point S-matrices completely fixed once the $ 4 $-point S-matrix is known. Since there is particle production, we will have non-trivial higher-point S-matrices as well. The next S-matrix that should be evaluated for the $\phi^4$ theory considered here is the $ 6 $-point scattering. However, given the number of interaction vertices and the addition of two external legs, the diagrammatics becomes cumbersome. Even if one perseveres and works out the $ 3 \rightarrow 3 $ amplitude, going to higher points would be prohibitively difficult. We entertain the hope that it may be possible to deploy recent advances in the computation of scattering amplitudes. In particular, the perturbiner method \citep{Rosly:1996vr,Mafra:2015gia,Mizera:2018jbh,Gomez:2020vat} guarantees a Berends-Giele-like recursion relation for \emph{any} quantum field theory and has recently been extended to $ 1 $-loop level as well \cite{Lee:2022aiu,Upcoming:2022}. A successful adaptation of the perturbiner method to evaluate the perturbative S-matrices in these theories will hopefully reveal structures the CDD factor imposes on renormalisability of non-integrable quantum field theories. 

A second follow-up question involves the supersymmetric version of $\ttb$ deformations. There are analogous deformations which are known to preserve both $\mathcal{N}=1$ and $\mathcal{N}=2$ SUSY \cite{Baggio:2018rpv,Chang:2018dge,Chang:2019kiu,Ferko:2019oyv,Jiang:2019hux,Cribiori:2019xzp,Ebert:2020tuy}. One would naturally expect an analogue of CDD factor to exist for S-matrices of these deformed theories as well. The existence of integrability and supersymmetry should impose very strong constraints on the quantum dynamics and it is possible that nonrenormalisation theorems can be systematically proved for these theories. Even if the undeformed theory is non-integrable, given the indications we have uncovered in this paper, one can reasonably expect some definite structure to exist dictated by an intriguing interplay of supersymmetry, integrability, and renormalisation.

We leave the exploration of these problems for the future.

\acknowledgments{We would like to thank Divyanshu Gupta for collaboration at an earlier stage of this project. SC would like to thank Renann Lipinski Jusinskas for numerous discussions and suggestions as well as for sharing the preliminary results of \cite{Upcoming:2022}. Research of SC has been supported by the Czech Science Foundation - GAČR, project 19-06342Y. MR is supported by Grant No. 21/02253-0 and 19/21281-4, São Paulo Research Foundation (FAPESP).}

\appendix

\section{Conventions}
\label{app:conventions}

We collect various conventions we have used throughout the paper in this appendix.

We work in flat spacetime with signature $(-,+)$, and for discussing
self-duality we will need the Levi-Civita symbol $\epsilon^{\mu\nu}$
with conventions $\epsilon^{01}=-\epsilon^{10}=1$.

Two-dimensional Fourier transforms are defined as
\begin{equation}
	\begin{aligned}
		f(x) &= \int \frac{\dd^{2} k}{(2\pi)^{2}} \, e^{ikx} \,\tilde{f}(k) \ , \\
		\tilde{f}(k) &= \int \mathrm{d} ^{2} x \, e^{-ikx} \, f(x) \ .
	\end{aligned}
\end{equation}

Since we have just a single independent component in the self-dual
form field $ A_{\mu} $, it is useful to work in lightcone
coordinates. Let us define the following lightcone coordinates and
derivatives:
\begin{equation}
	X^{\pm} = \frac{X^{1} \pm X^{0}}{\sqrt{2}} \quad \text{and}
	\quad \partial_{\pm} = \frac{1}{\sqrt{2}} \left( \partial_{1}
	\pm \partial_{0} \right) \ ,
\end{equation}
and inversely,
\begin{equation}
	X^{0/1} = \frac{1}{\sqrt{2}} \left( X^{+} \mp X^{-} \right)
	\quad \text{and} \quad \partial_{0/1} = \frac{1}{\sqrt{2}}
	\left( \partial_{+} \mp \partial_{-} \right) \ .
\end{equation}

In these coordinates, the metric is $\eta_{-+} = \eta_{+-} = 1$ such
that $\eta = \eta^{-1}$. Correspondingly, the Levi-Civita symbol is
now $\epsilon_{-+} = -\epsilon_{+-} = -\epsilon^{-+} = \epsilon^{+-} =
1$. Now, the self-dual gauge field with components $A_{\mu} = (-a,a)$
in Cartesian coordinates is such that its index can be raised and
lowered using the metric, but it also obeys a self-duality condition
$A^{\mu}= \epsilon^{\mu \nu}A_{\nu}$. Demanding that these two
conditions are compatible yields that
\begin{equation}
	A_{+} = A^{-} = 0 \quad \text{and} \quad A_{-} = A^{+} \neq 0
	\ .
\end{equation}
We fix the non-zero components by computing an inner product, say
$X\cdot A$, in two coordinate systems and demanding that they
agree. This allows us to conclude that $A_{-} = A^{+} =
\sqrt{2}a$. Similarly, for an anti-self-dual field with components
$B_{\mu} = (b,b)$ in Cartesian coordinates will have $B_{+} = B^{-} =
-\sqrt{2}b$.

As for the projectors, in lightcone coordinates we have that
\begin{equation}\label{key}
	\left(\mathcal{P}_{+}\right)^{+-} =
	\left(\mathcal{P}_{+}\right)_{-+} =
	\left(\mathcal{P}_{-}\right)^{-+} =
	\left(\mathcal{P}_{-}\right)_{+-} = 1 \ ,
\end{equation}
and the other possibilities identically zero. This is good: it tells
us that only $ A_{-} $ and $ B_{+} $ propagate, which makes sense
because they are chiral.

Since we're working in lightcone coordinates, for massless particles
we will have $p^{2}=2p_{+}p_{-}=0$ so either $p_{+/-}=0$. For the case
of $ 2\rightarrow 2 $ scattering, we make the choice that particles
labelled $1,3$ will have $p_{+}=0$ and for particles labelled $2,4$ we
have $p_{-}=0$. This ensures that $s=(p_{1}+p_{2})^{2} = 2p_{1}\cdot
p_{2} = 2\left(p_{1}\right)_{-} \left(p_{2}\right)_{+} = -2 p_{1}\cdot
p_{4} = -t$, and that $u = (p_{1}+p_{3})^2=0$. These conventions make
sense because $ A_{\mu} $ has non-zero component $ A_{-} $ and that
should have non-zero momentum, and similarly with $ B_{\mu} $ and $
B_{+} $.

\section{Feynman Rules for Chiral Bosons} \label{app:feynmanrules}
Here we give the explicit Feynman rules for the chiral boson theory in terms of the original field variables following Sen's formalism. We derive the Feynman rules in such a way that it can be easily adapted for case of higher self-dual $p$-form field strengths as well.  

The action for a pair of left- and right-chiral bosons interacting via linear deformation in the $\ttb$ coupling is given by
\begin{align}
\mathcal{L}= \frac{1}{2} \Big( \partial _\mu \phi \partial ^\mu \phi +  \partial _\mu \tilde{\phi }\partial ^\mu \tilde{\phi} \Big) -2\epsilon ^{\mu \nu}\Big(A_\mu \partial _\nu \phi -B_\mu \partial _\nu \tilde{\phi} \Big) +\frac{\lambda}{4} \Big(A_\mu B^\mu \Big)^2 \,,
\end{align}
where $\phi, \tilde{\phi}$ are the auxiliary scalars needed to describe chiral 1-form fields in Sen's formalism. 

Since the derivation of the propagators (in momentum space) have already been discussed in the main text, we simply quote them here for completeness
\begin{align}
	\langle A_\mu(-k) A_\nu(k) \rangle &=-\frac{i}{4} \frac{k_\mu k_\nu}{k^2} \,;\\
	\langle B_\mu(-k) B_\nu(k) \rangle &=-\frac{i}{4} \frac{k_\mu k_\nu}{k^2} \;.
\end{align}
Note that it is implicit that the propagators are always sandwiched between appropriate projectors as pointed out in \citep{Sen:2015nph}. In particular, note that there are no contractions between $A_\mu$ and $B_\nu$.

 The vertex factor can be read of from the action as usual. There is only one additional subtlety: the vertex must keep the (anti-)self-duality of each leg manifest. Recall that in Sen's formalism, the (anti-)self-duality holds off-shell as well, so this is a crucial and necessary step. Ensuring the correct duality property of each leg in a vertex is easily taken care of by inserting appropriate factors of the projectors. For the specific interaction term under consideration the vertex factor is
\begin{align}
\widehat{V}_{A_{\mu}B_\nu A_\alpha B_\beta } =V^{(4)}_{A_{\gamma}B_\delta A_\rho B_\sigma } \mathcal{P}^\rho{}_{-,\alpha}\mathcal{P}^\sigma{}_{+,\beta}\mathcal{P}^\gamma{}_{-,\mu}\mathcal{P}^\delta{}_{+,\nu} \ ,
\end{align}
where
\begin{align}
 V^{(4)}_{A_{\mu}B_\nu A_\rho B_\sigma } = \frac{i\lambda}{2} \big( \eta _{\mu \nu} \eta _{\rho \sigma} +\eta _{\mu \sigma} \eta _{\rho \nu}\big)\,,
\end{align}
is the vertex factor that one would have written if $A$ and $B$ were ordinary vector fields.
This vertex factor ensures that only chiral or anti-chiral particles are created or annihilated at the vertices, even off-shell.

Finally, since the field variables here are like field strengths, each external leg carries an additional factor of $\sim i k_\mu$ (with appropriate signs depending on the momentum flowing in or out as per a chosen convention). For most cases, we actually know the action in terms of potential as well the field strength. This indicates that once the wave function normalization is fixed for the potential, that automatically fixes the normalisation for the wave function of the field strength.

As an example, for ordinary scalars, if one adopts the normalisation that $\langle 0 | \varphi(x)|k\rangle =e^{i kx}$, then for the field strength $V_\mu = \partial_\mu \varphi$ we automatically know the wave function normalisation. In Sen's formalism there is no such way to fix the numerical factor for the external legs uniquely since the mapping between the self-dual field strength and its potential is not uniquely known. This is however not a problem, since with any choice of normalisation, the LSZ reduction guarantees that the S-matrix is independent of any such wavefunction renormalisation.

This Feynman rules are consistent with the ones prescribed in \cite{Alvarez-Gaume:1983ihn} (albeit there it was given for self-dual $5$-forms in $10$ dimensions). The difference is now they can be systematically derived from an action following the procedure outlined in \cite{Sen:2015nph} and we have given an explicit derivation for a concrete example to supplement the general procedure. Of course, the end result of a computation does not depend on the choice of field variables and for theories with equal number of chiral and anti-chiral fields it might be simpler to work with the fields $S_\mu$ defined in the text. However, one can easily imagine theories which break chiral symmetry or where only a self-dual field strength couples to other matter fields, for those cases it is unavoidable that one works with the Feynman rules that can be derived in the same spirit as it has been done in this appendix.

\section{Useful Integrals} \label{app:integrals}
In this appendix, we collect all the integrals used in \Secref{1loopphi4} of 1-loop amplitude for massive $\phi ^4$ theory following \cite{Rosenhaus:2019utc}. We begin with the simplest integral that does not have any tensor structure in the integrand
\begin{align}
L:=\int \frac{\mathrm{d}^2k}{(2\pi)^2}\frac{1}{k^2-m^2}\frac{1}{(k+P)^2-m^2}=\frac{i}{4\pi} \frac{(i\pi -\theta)}{m^2 \sinh \theta}\,,
\end{align}
where $P:=p_1+p_2 $ and $\theta:= \theta _1-\theta _2 $,  relative rapidity between particles with momentum $p_1$ and $p_2$.   This result has been used in \eqref{noderivative} giving the contribution of no derivative terms in 1-loop amplitude.  Next we consider the following symmetric second rank 1-loop integral
\begin{align}
L_{\mu \nu}:=\frac{1}{4} \int \frac{\mathrm{d}^2k}{(2\pi)^2}\frac{k_\mu}{k^2-m^2}\frac{P_\nu-k_\nu}{(k+P)^2-m^2}\,.
\end{align}
These class of integrals are required in the evaluation of the single derivative terms appearing in the integrand of massive $\phi ^4$ 1-loop amplitude.
In lightcone coordinates,  the components of this tensor integral are given by
\begin{align}
L_{++} &=-\frac{i}{16 \pi}e^{(\theta _1+\theta _2)} \left(1-\frac{(i\pi -\theta)}{\sinh \theta} \right)\,,\\
L_{--} &=-\frac{i}{16 \pi}e^{-(\theta _1+\theta _2)} \left(1-\frac{(i\pi -\theta)}{\sinh \theta} \right)\,,\\
L_{-+} &=-\frac{i}{16 \pi} \left(\log \left(\frac{m^2}{\Lambda ^2}\right)-(i\pi -\theta)\coth \theta \right)\,.
\end{align} 
Finally, we consider a class of rank-4 tensor integrals, appearing as double derivative terms in the 1-loop integrand of \eqref{phi4integrand} 
\begin{align}
L_{\mu \nu \rho \sigma}:=\frac{1}{16} \int \frac{d^2k}{(2\pi)^2}\frac{k_\mu k_\nu}{k^2-m^2}\frac{\left(P_\rho-k_\rho \right)\left(P_\sigma -k_\sigma \right)}{(k+P)^2-m^2}\,.
\end{align}
By definition, this tensor integral is symmetric within the first two and last two indices and also between the pairs.  Again,  in lightcone coordinate, the different components of this tensor integral has been evaluated in \cite{Rosenhaus:2019utc} and we quote the results below for completeness
\begin{align}
L_{++++} &=-\frac{im^2}{64 \pi}e^{2(\theta _1+\theta _2)} \left( \frac{4}{3} + \frac{\cosh \theta}{3} - \frac{ ( i \pi\, - \theta)}{ \sinh \theta}\right)\\
L_{-+++} &= \frac{im^2}{192 \pi} e^{(\theta _1+\theta _2)} \Big( 1 - 2\cosh\theta + 3(i\pi - \theta)\frac{\cosh \theta}{\sinh \theta}\Big) \\
L_{- - + + } &=\frac{i m^2}{64\pi} \left(\frac{\Lambda ^2}{m^2}-2 \cosh\theta  \log \frac{m^2}{\Lambda ^2}+(i \pi -\theta)\frac{ \cosh (2\theta)}{\sinh \theta}\right)\\
L_{+ - + -} &=\frac{i m^2}{64\pi} \left(\frac{\Lambda^2}{m^2} - \frac{1}{3} \cosh \theta + 2 \log \frac{m^2}{\Lambda ^2} + \frac{(i \pi - \theta)}{\sinh \theta} \right)\,.
\end{align} 
All other components of the rank-4 tensor integral are related to the above integrals. For instance:
\begin{align}
L_{----}=e^{-4(\theta _1+\theta _2)}L_{++++}\,, \quad L_{+---}=e^{-2(\theta _1+\theta _2)}L_{-+++}\,.
\end{align} 
Symmetry considerations also constrains some of the integrals as follows
\begin{align}
L_{++--}=L_{--++}\,, \quad L_{+-++}=L_{-+++}\,,
\end{align}
and so on. These results have been used to determine the contributions coming from single derivative terms ($\mathcal{A}^{(b)}$ in \eqref{singlederivative}) and double derivative terms ($\mathcal{A}^{(c)}$ in \eqref{doublederivative}) in massive $\phi ^4$ 1-loop amplitude.

\bibliographystyle{JHEP}
\bibliography{Refs}
% End Document ----------------------------------------------------------------
\end{document}